\documentclass{article}

\usepackage[final, nonatbib]{neurips_2021}

\usepackage{float}

\usepackage[utf8]{inputenc} 
\usepackage[T1]{fontenc}    
\usepackage{hyperref}       
\usepackage{url}            
\usepackage{booktabs}       
\usepackage{amsfonts}       
\usepackage{nicefrac}       
\usepackage{microtype}      
\usepackage{xcolor}         
\usepackage{amsmath,amsfonts,amsthm,bm}
\usepackage[pdftex]{graphicx}
\usepackage[english]{babel}
\usepackage{blindtext}

\title{Double U-Net for Super-Resolution and Segmentation of Live Cell Images}

\author{%
  Mayur Bhandary, J. Patricio Reyes, Eylül Ertay, Aman Panda  \\
  Cornell Tech \\
}
\begin{document}

\maketitle

\begin{abstract}
Accurate segmentation of live cell images has broad applications in clinical and research contexts. Deep learning methods have been able to perform cell segmentations with high accuracy; however developing machine learning models to do this requires access to high fidelity images of live cells. This is often not available due to resource constraints like limited accessibility to high performance microscopes or due to the nature of the studied organisms. Segmentation on low resolution images of live cells is a difficult task. This paper proposes a method to perform live cell segmentation with low resolution images by performing super-resolution as a pre-processing step in the segmentation pipeline. 
\end{abstract}

\section{Introduction}

Image segmentation is a broadly applicable practice within computer vision that has recently been revolutionized through deep learning methods. Image segmentation is the process of identifying pixels that belong to the same class of object. In practice, deep learning approaches with U-Net model architectures present a very effective way to perform image segmentation. This architecture contains a contracting path and symmetric expanding path for the detection of context and localization respectively, and a network training strategy based on data augmentation \cite{ronneberger2015u}. In the PhC-U373 and DIC-HeLa datasets, in the task of cell segmentation in light microscopic images, the U-Net achieved an average intersection over union (IoU) much higher than any past segmentation results \cite{ronneberger2015u}, defining the state-of-the-art for image segmentation.

To achieve high segmentation accuracy, the input images require a minimum resolution. However, in various contexts high resolution imaging is often not viable due to the limitations of the imaging equipment or the nature of the captured phenomena. Deep learning presents opportunities to enhance images with techniques such as super-resolution. Super-resolution is a process used to estimate a high resolution image based of a low resolution input image.\cite{romano2016raisr}

By applying super-resolution pre-processing to the images prior to feeding them into a segmentation network it is possible to increase the quality of the final segmentation. In fact, this approach has been successfully applied in the biological imaging context, where high resolution inputs are often not available.
For example, in the magnetic resonance imaging (MRI) domain,  high resolution imaging is possible, but requires a generous scanning time and covers less spatial area per unit time when detail is prioritized \cite{chen2018efficient}. Super-resolution offers a means by which low resolution MRI can be transformed into high resolution MRI \cite{chen2018efficient}, sidestepping the constraints mentioned.

This paper proposes a new approach to work with low resolution cell images by performing a double U-net, starting with a super-resolution network on a downsampled, low quality image cell dataset, then performing segmentation. We show that using a double structured U-Net improves the segmentation results.

\section{Related Work}

\subsection{Super-resolution using Deep Learning on Cell Images}

Since the acquisition of high quality images is not always feasible in many settings, particularly in medicine, the subject of super-resolution has been trialed and researched very often. Super-resolution is abundantly used for live cell imaging, especially in cell images gathered via techniques such as multi-photon microscopy and structured illumination microscopy \cite{li2020adaptive, zhou2020w2s}.  Wang et al. \cite{wang2019deep} trained a generative adversarial network to transform diffraction-limited input images of cells gathered using fluorescence microscopy into super-resolved images. Furthermore, Qiao et al. \cite{qiao2021evaluation} deployed and trained a deep Fourier channel attention network (DFCAN) to obtain super-resolution images of diverse biological structures, where their model achieved comparable image quality to images obtained with higher resolution. In Hu et al. \cite{hu2019runet}, a general purpose super-resolution architecture based on U-Net, dubbed the RUNet is proposed, which shows the accurate production of high quality images from lower quality images in a variety of domains.

\subsection{Segmentation of Cells}

Detection and tracking of individual cells within cell groups is a widely used and important practice in cancer research, and other areas in medicine such as drug discovery \cite{al2018deep}. Hatipoglu et al. \cite{hatipoglu2017cell} compared three deep learning algorithms, namely Convolutional Neural Networks, Deep Belief Networks and Sparse Autoencoders in cell segmentation, finding that the CNN outperformed other deep learning algorithms. 

In their research aiming to mitigate the issues with having low resolution images for image segmentation, Fang et al. \cite{fang2021deep} applied a technique of using a "crappifier" to downsample high resolution images using a Res-Net based U-Net, similar to the proposed work of our project. As far as our literature review goes, using a double U-net based pipeline to apply super-resolution from downsampled high-resolution images which are then used later on for segmentation has not been discovered in previous research projects.

There are many accuracy measures used to gauge the performance of segmentation networks. Edlund et al. \cite{edlund2021livecell} used 3 measures of accuracy for segmentation in their research; namely average precision (AP), average false-negative ratio (AFNR), and intersection over union (IoU). Previous literature shows that in binary segmentation, Dice score is widely used due to its robustness to class imbalance \cite{fidon2017generalised, bertels2019optimizing, zhang2017brain}.

\subsection{U-Net based models}

U-Net is a model initially proposed by Ronneberger et al. \cite{ronneberger2015u} that won the ISBI cell tracking challenge. The model is based on a CNN architecture with 2 paths, one for up-sampling and one for downsampling. The structure of the model is as shown in the following figure.

There have been modified implementations of the U-Net, such as the MultiResUNet \cite{ibtehaz2020multiresunet} where the two convolutional layers are replaced by "MultiRes blocks" which consist of 3×3, 5×5 and 7×7 convolutional filters in parallel, and U-Net++ that consists of encoder-decoder sub-networks, designed to skip pathways \cite{zhou2018unet++}.

\begin{figure}[H]
\centering
\includegraphics[width=0.9\linewidth]{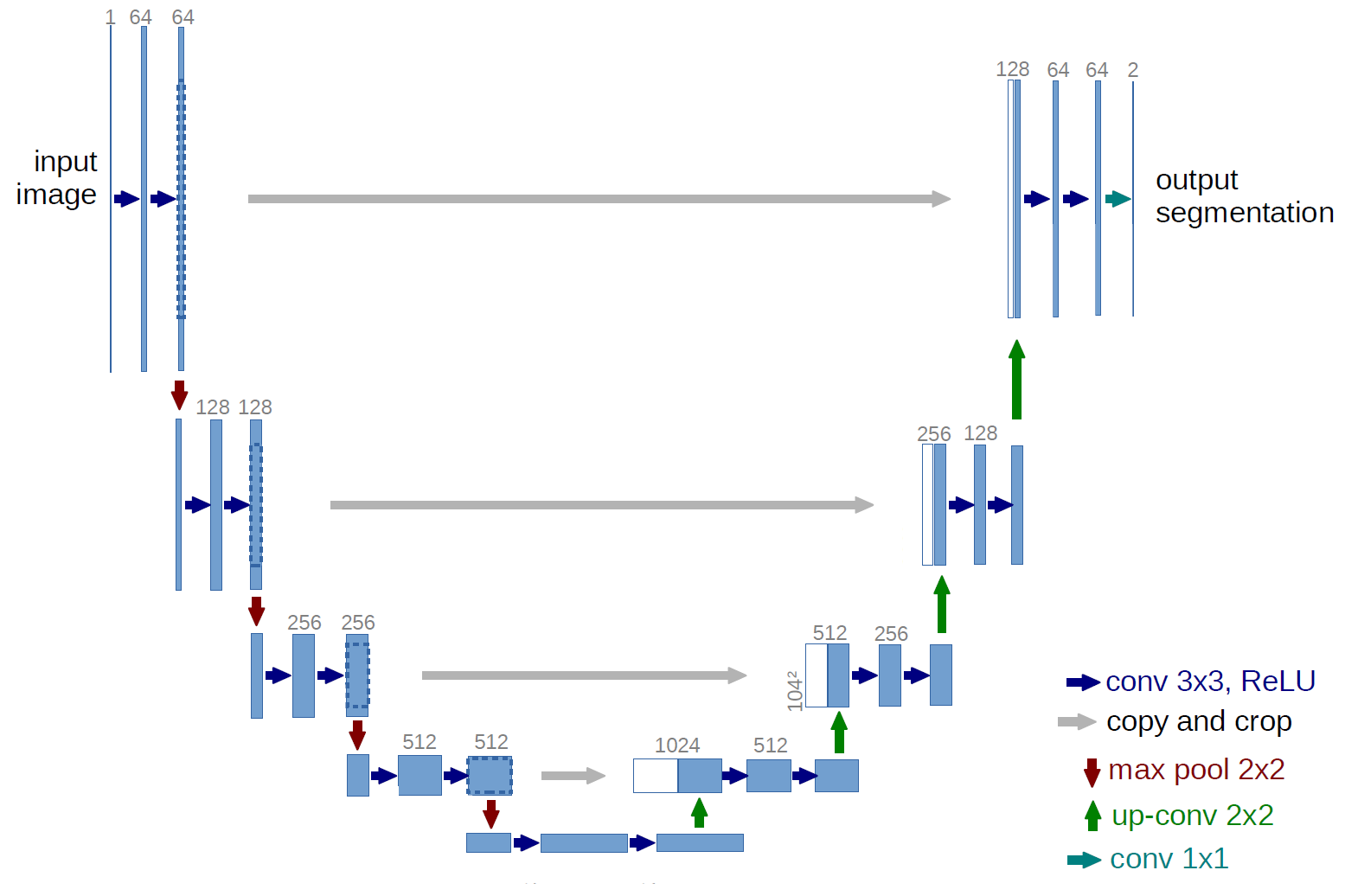}
\caption{The U-Net Model architecture}
\end{figure}

\section{Method}

The proposed implementation is broken up into two main phases. The first phase requires training a U-Net for segmentation and the second phase requires training a U-Net for super-resolution. In the first phase, high resolution images are paired with their corresponding masks to train the segmentation U-Net. After training, performance of the segmentation model is evaluated on both the original high resolution test dataset and on downsampled versions. In the second phase, the super-resolution network is trained using low resolution images and their high resolution correspondances. Finally, the segmentation model is evaluated again using the low resolution dataset but this time with the super-resolution model a pre-processing step. 

\begin{figure}[H]
\includegraphics[width=\linewidth]{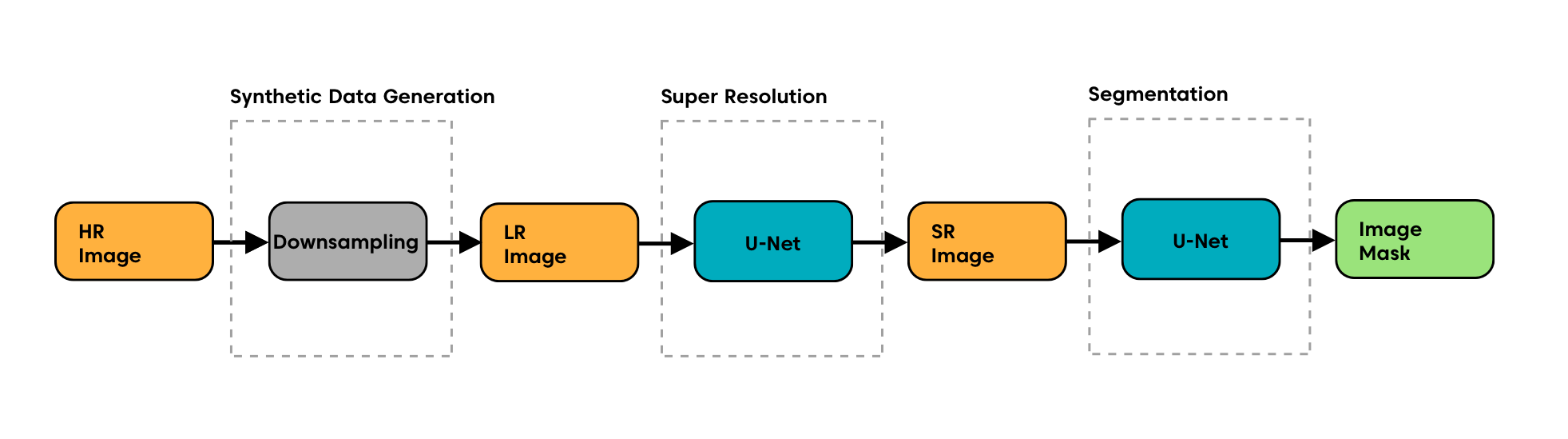}
\caption{The Double U-net pipeline}
\end{figure}

\begin{figure}[H]
\includegraphics[width=\linewidth]{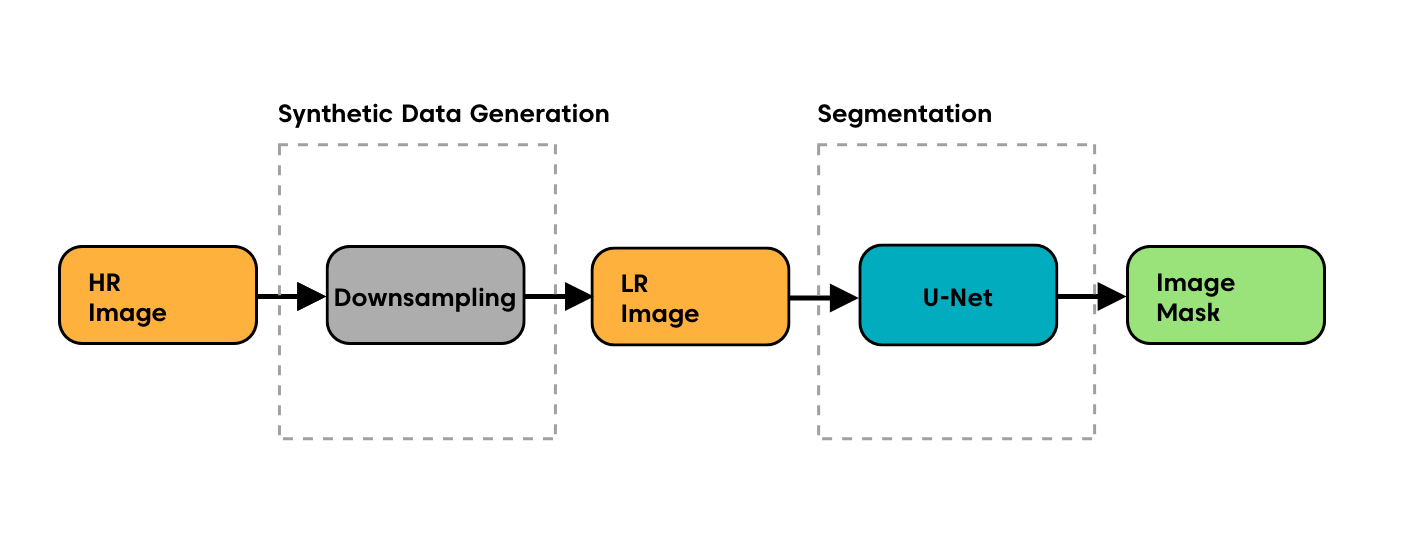}
\caption{Baseline segmentation pipeline}
\end{figure}

\subsection{LIVECell Dataset}

We use the LIVECell dataset from Edlund et al.'s paper, "LIVECell—A large-scale dataset for label-free live cell segmentation"\cite{edlund2021livecell}. This dataset contains images of over 1.6 million cells obtained via phase contrast microscopy. Images of 8 different types of cells are included in this data set which represent a variety of cell body types. The quality of our segmentations are evaluated using the Sørensen–Dice coefficient (Dice Score) \cite{bertels2019optimizing}. This data uses Microsoft's COCO format to specify segmentation annotations. The dataset provides annotations for each individual cell in the frame. In order to create the ground truth mask for each image, we superimposed each cell annotation and converted the resulting image into a binary mask. 

\subsection{Downsampling}

In order to train the super-resolution network, we first generated degraded versions of our input images by downsampling them. Downsampling can be used to model the  limitations of physical systems in generating high resolution images. For example, in live cell imaging via multi-photon microscopy, there is a limit to the amount of power that can be applied to cells before they become damaged \cite{li2020adaptive}. Downsampling is a reasonable transformation to mimic the reduced resolution of this type of imaging because photons are applied at discrete intervals of the sample in order to limit the power delivered to the cells. The degraded images are the inputs to our super-resolution network and the corresponding original images are the targets.

We do this by applying a downsampling factor, and then up-sampling by the reciprocal. We specifically use downsampling factors of 0.25 and 0.125. For the former, this takes the dimension of the original high resolution images from 704 x 520 to 176 x 130, and then back to 704 x 520 and for the latter, 704 x 520 to 88 x 65, and then back to 704 x 520.

\subsection{Double U-Net}

For the two U-Net models we started with a PyTorch implementation of the U-Net for image segmentation, originally used for Kaggle's Carvana Image Masking Challenge\cite{milesi}.

\begin{figure}[H]
\includegraphics[width=0.9\linewidth]{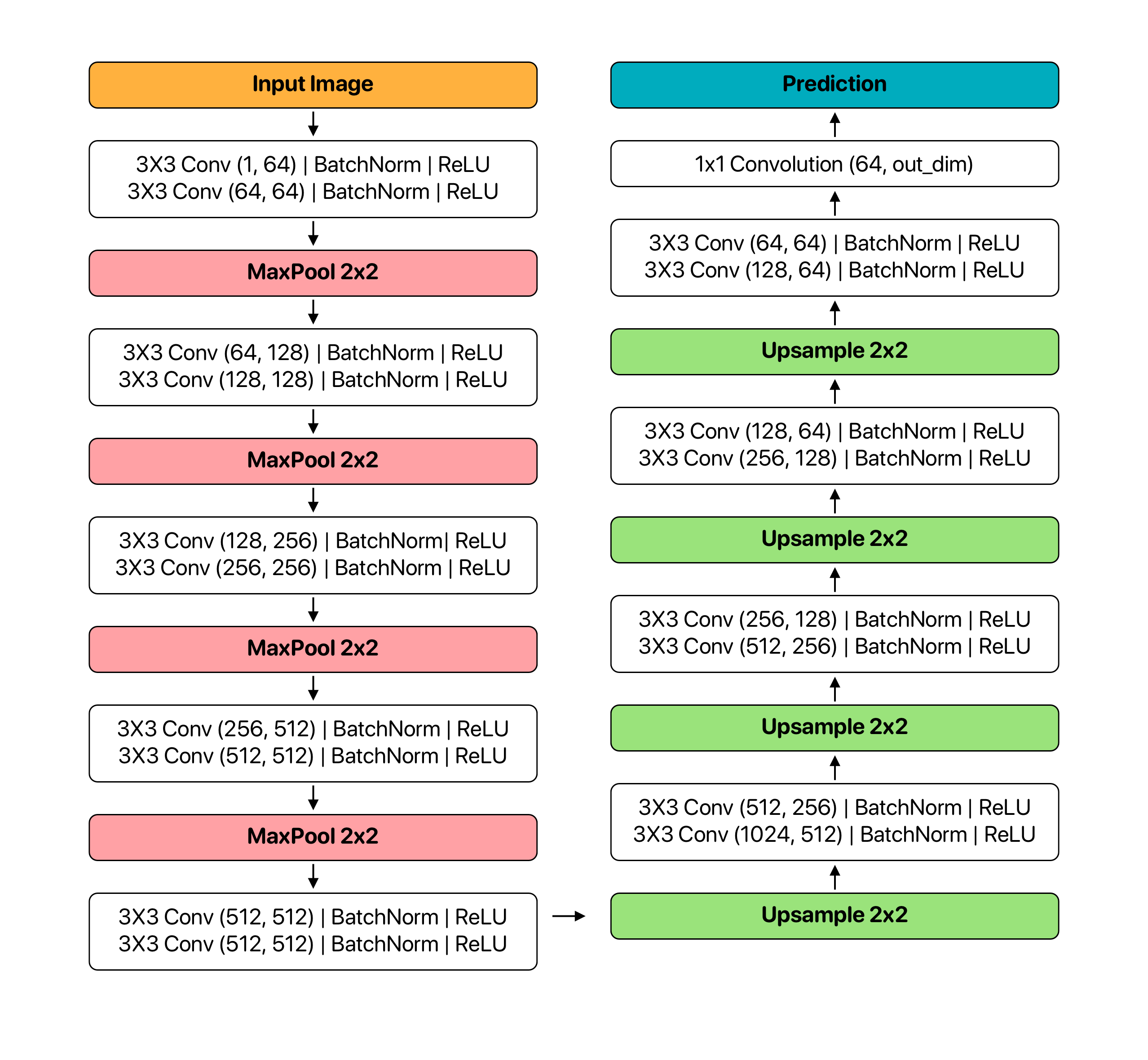}
\caption{The U-Net implementation}
\label{fig:unetdetail}
\end{figure}

As seen on figure \ref{fig:unetdetail}, this model uses the U-Net architecture, consisting mainly of 3X3 convolutional layers followed by ReLU activation functions and MaxPool/Upsampling layers. Additionally, it incorporates batch normalization after each 3x3 convolutional layer. All 3x3 convolutions are performed with a padding of 1 and stride of 1. The output dimension for the segmentation model is the number of classes (2); background and cell. The same architecture was used for super-resolution by changing the output dimension to 1.

\subsubsection{Super-resolution}

The implementation of the proposed model required modification to perform super-resolution rather than segmentation. This was achieved by changing the output layer dimension from 2 to 1 and removing the final Softmax layer. Mean squared error was used as the objective function for super-resolution, which is given by

\[MSE = \frac{1}{n}\sum_{i = 1}^{n}(Y_i - \hat{Y})^2\]

\subsubsection{Cell Segmentation}

Apart from minor tweaks to work with our novel dataset, the existing U-Net implementation did not require many changes. In the experimentation phase, cross entropy loss was used as the objective function for this model. The cross entropy loss is given by

\[H(p,q) = -\sum p(x) log(q(x)) 
\] \cite{ronneberger2015u}.

\subsection{Accuracy}

The Dice score is a commonly used metric to evaluate the quality of image segmentation in a medical context \cite{bertels2019optimizing, zhang2017brain, bertels2019optimizing}. As a consequence, it was our metric of choice to measure accuracy from our second image segmentation U-Net. Intuitively, it is a statistic that measures the similarly between 2 segmentation masks in our context. It is given by the following, where |X| and |Y| are the cardinalities of the generated  
mask and the ground truth mask respectively.

\[\frac{2|X\cap Y|}{|X| + |Y|}\]

\section{Results}

\subsection{Establishing the Baseline}

To evaluate the merits of the proposed double U-Net architecture, the following boundaries were established: On the upper bound, there is the Dice score of the segmentation U-Net running directly on the high resolution LIVECell images. On the lower bound, there is the Dice score of the segmentation U-Net running directly on the downsampled LIVECell images to emulate segmentation performance given resource constraints. Table \ref{table:1} shows the results of the Dice scores, showing that the high resolution images achieve a very impressive Dice Score of $0.9$ on both test and validation sets, and the downsampled images perform poorly $(0.3)$. The hypothesis evaluated in this study is that performing super-resolution on the downsampled LIVECell images to subsequently aid in the segmentation U-Net in producing the masks, producing a higher Dice score compared to the baseline of getting the segmentation from the downsampled images direclty. Firstly, the segmentation U-Net was trained on the training set of the LIVECell data set. The results of the trained U-net on the high-resolution LIVECell test images are presented below, as well as the results of generated images and Dice scores below for the downsampled test set (with scale factors of 0.25 and 0.125) as a baseline to compare the super-resolved downsampled images.

\begin{table}[h]
\centering
\begin{tabular}{||c c c||} 
 \hline
 Image & Test Set Dice Score & Validation Set Dice Score \\ [0.5ex] 
 \hline\hline
 High resolution, original image &  0.906 & 0.897 \\ 
 Downsampled image, M = 0.25 & 0.317 & 0.358 \\
 Downsampled image, M = 0.125 & 0.258 & 0.297 \\ [1ex] 
 \hline
\end{tabular}
\caption{Table to test captions and labels.}
\label{table:1}
\end{table}

\begin{figure}[h!]
\includegraphics[width=\linewidth]{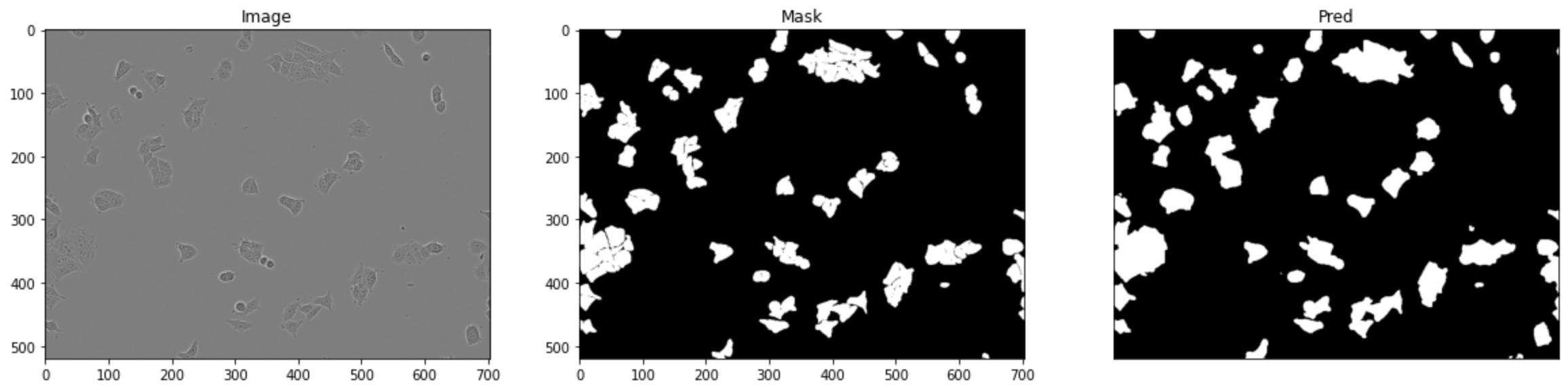}
\caption{Results of the Segmentation U-net on the Test set}
\label{fig:segmentation-actual}
\end{figure}

\begin{figure}[h!]
\includegraphics[width=\linewidth]{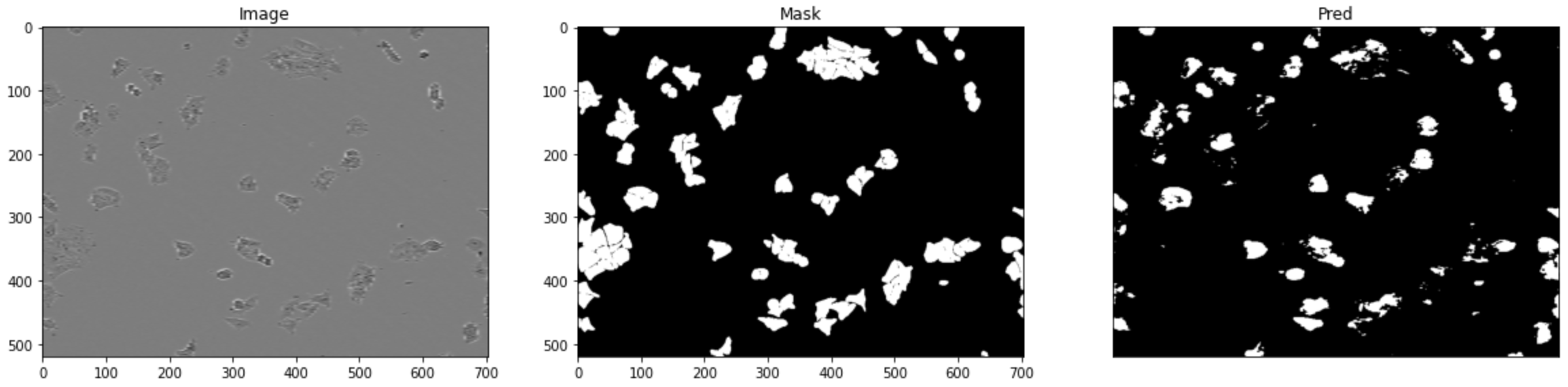}
\caption{Results of the Segmentation U-net on the Downsampled Image with downsampling rate $M = 0.25$}
\label{fig:segmentation-down1}
\end{figure}

\begin{figure}[h!]
\includegraphics[width=\linewidth]{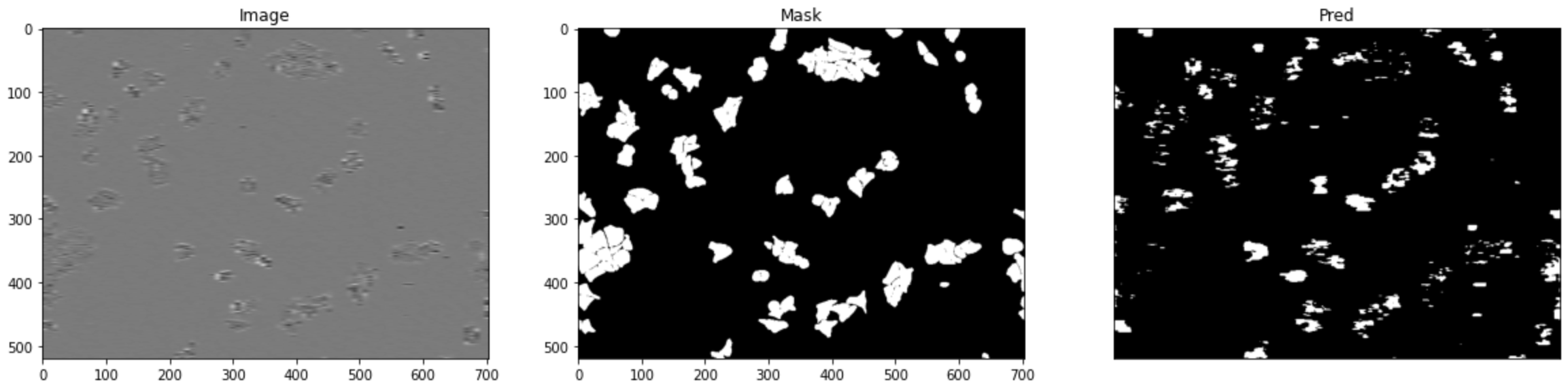}
\caption{Results of the Segmentation U-net on the Downsampled Image with downsampling rate $M = 0.125$}
\label{fig:segmentation-down2}
\end{figure}

\newpage

\subsection{Segmentation on Super-resolution}

\begin{figure}[h]
\includegraphics[width=\linewidth]{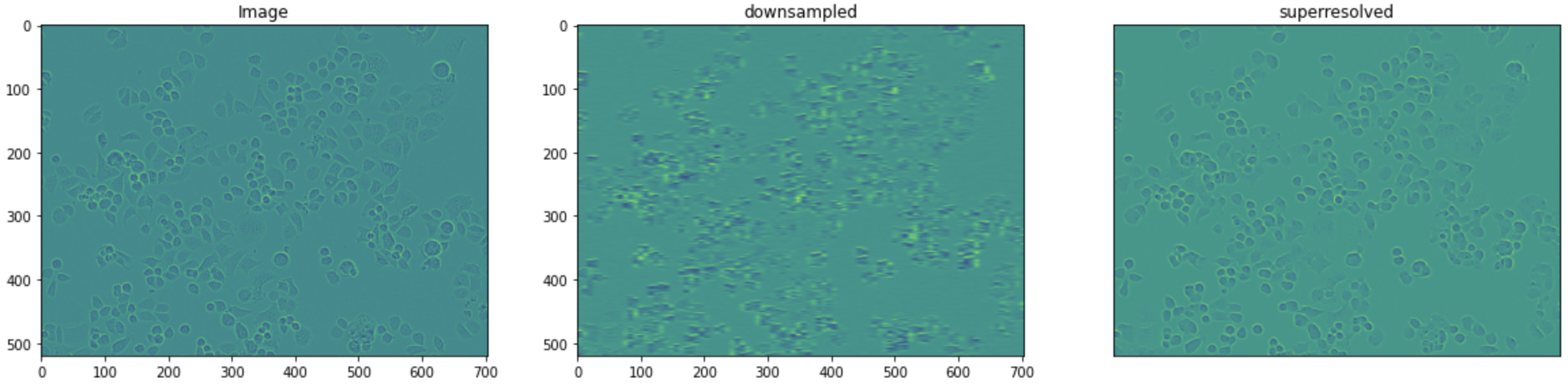}
\caption{Results of the Super-resolution U-net on the Downsampled Image with downsampling rate $M = 0.125$}
\label{fig:sup}
\end{figure}

We trained a super-resolution U-Net each for the downsampled LIVECell training images with 0.25 and 0.125 factors, using the corresponding high resolution images as the ground truth for each model. We then ran the downsampled test set images through the super-resolution U-Nets to get super resolved images. Finally, we input the two super resolved test image sets through the trained segmentation U-Net to get the following results. The results can be seen in table \ref{table:2}. The output image of the super-resolution U-net can be seen in figure \ref{fig:sup}.

\begin{table}[h]
\centering
\begin{tabular}{||c c c||} 
 \hline
 Image & Test Set Dice Score & Validation Set Dice Score \\ [0.5ex] 
 \hline\hline
 Downsampled image, M = 0.25 & 0.384 &  0.437 \\
 Downsampled image, M = 0.125 & 0.385 & 0.452 \\ [1ex] 
 \hline
\end{tabular}
\caption{Evaluation after Performing Super-Resolution}
\label{table:2}
\end{table}

\begin{figure}[h!]
\includegraphics[width=\linewidth]{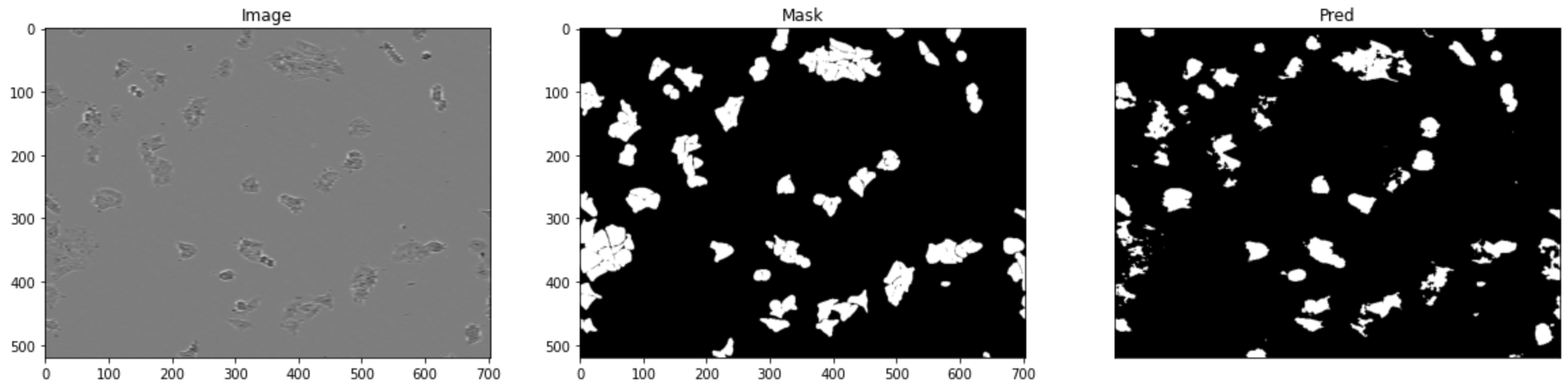}
\caption{Results of the Segmentation U-net Super-resolved, Downsampled image with downsampling rate $M = 0.25$}
\label{fig:segmentation-1}
\end{figure}

\begin{figure}[h!]
\includegraphics[width=\linewidth]{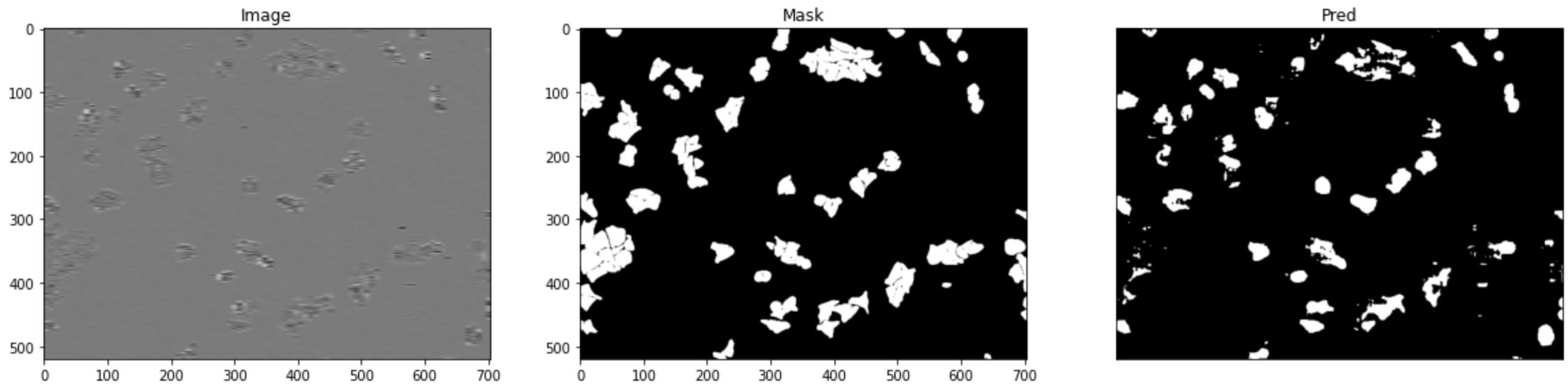}
\caption{Results of the Segmentation U-net Super-resolved, Downsampled image with downsampling rate $M = 0.125$}
\label{fig:segmentation-2}
\end{figure}

\newpage

\subsection{Discussion}

Super-resolution on the downsampled images produced a marked increase in Dice score from the second U-Net. Namely, Dice score increased by about 7\% for the higher resolution downsampled images, and 13\% for the lower resolution downsampled images. However, an interesting result was the test set Dice score for both downsampling levels resulting in an almost equivalent accuracy of 38\%. This seems to imply that there is a ceiling in the improvement super-resolution can provide for segmentation accuracy. This can be explained by details that are lost in the downsampling process that can not be hallucinated accurately by the super-resolution U-Net.

On initial observation, the super-resolved images produced by the super-resolution U-Net are impressively detailed and seem close to interchangeable with the actual high resolution images. However, on further inspection, one can see that the super-resolved images appear smoother with less pronounced detail compared to the original high resolution image. This effect proves as a bottleneck to the subsequent segmentation U-Net as the masks that are generated seem to be more conservative than the ground truth masks. This trend can be observed by comparing figures \ref{fig:segmentation-down1} and \ref{fig:segmentation-down2} to \ref{fig:segmentation-1} and \ref{fig:segmentation-2}.

Upon inspecting the Dice scores on downsampled images before and after super-resolution, the hypothesis that a custom double U-net implementation would significantly improve the results of segmentation has been verified. 

\section{Conclusion}

We propose a novel approach to use super-resolution for low quality cell images, due to imaging techniques or the fragility of live cells. We use a recently published dataset, LIVECell, that has been trained on CNN-based networks only for segmentation. The practice of applying simultaneous super-resolution and segmentation has been explored in other contexts; whereas this paper demonstrates the validity in applying a similar methodology in segmenting live cell images gathered via phase contrast microscopy. The results of the experiments showed that using a double U-net can improve the segmentation accuracy of masks in contexts where high quality imaging is not feasible. We propose this as a potential method for live-cell imaging microscopy, to construct a better guideline for shining light on the more accurately detected location of images, which is a problem often faced in light microscopy. 

\section{Future Work
}
In this paper, we verified that super-resolution allows the segmentation model to more accurately predict edges. These edges appear to be substantially obscured in the downsampled cell images. Comparatively, the majority of the cell surface area, which appears to be mostly vacant space, might not be detail lost in low resolution. As a natural continuation of this work, we will test this hypothesis by applying an edge detection filter to the input images and comparing the edge pixels to the coverage of the segmentations. For example, a high pass filter could be applied to extract cell body edges, and a thresholding operation could be used to create a binary mask for the edges. The same steps would then be applied to the predicted segmentations to generate the predicted edge masks. Edge masks produced by the model where super-resolution was applied first should have a higher Dice coefficient than masks produced without super-resolution. We hypothesize that this difference between low resolution and super-resolved images will be more pronounced than considering the cells at large.

Applying and then evaluating this double U-Net structure on different datasets would allow us to generalize this architecture to other contexts. Our motivation was the unavailability of high resolution live cell images due to microscope and other equipment limitations, as well as the fragility of some types of live cells. One other motivation is to optimize data storage when on-demand image segmentation is necessary, but we wish to store downsampled images to save on storage costs.

Another avenue for improvement would be to produce a single model that performs super-resolution and segmentation simultaneously. This could be implemented by summing the losses for segmentation and super-resolution to create an aggregate loss.

\section{Author Contributions}

All team members contributed to the literature review and writing the paper.

Mayur worked on data preprocessing and created the dataloader for generating masks and images to train our models. He also modified the U-Net for superresolution and trained the superresolution network. He also worked on adapting the Carvana code for segmentation.

J. Patricio worked on the code for Live Cell segmentation. He fine-tuned the parameters and trained the segmentation model. He also contributed on the code that combines both models into a Double U-Net architecture.

Eylül worked on the code for evaluating different models. She experimented with different accuracy metrics including pixel accuracy and dice coefficient. She worked on the code for the combined U-net architecture.

Aman worked on the data downsampling and upsampling procedure, experimenting with different factors that was ultimately used in the dataloader. He also worked on running the Live Cell implementation that was ultimately not used in lieu of the Carvana code implementation.

\clearpage

\bibliography{bibliography}

\begin{thebibliography}{10}
\providecommand{\url}[1]{#1}
\csname url@samestyle\endcsname
\providecommand{\newblock}{\relax}
\providecommand{\bibinfo}[2]{#2}
\providecommand{\BIBentrySTDinterwordspacing}{\spaceskip=0pt\relax}
\providecommand{\BIBentryALTinterwordstretchfactor}{4}
\providecommand{\BIBentryALTinterwordspacing}{\spaceskip=\fontdimen2\font plus
\BIBentryALTinterwordstretchfactor\fontdimen3\font minus
  \fontdimen4\font\relax}
\providecommand{\BIBforeignlanguage}[2]{{%
\expandafter\ifx\csname l@#1\endcsname\relax
\typeout{** WARNING: IEEEtran.bst: No hyphenation pattern has been}%
\typeout{** loaded for the language `#1'. Using the pattern for}%
\typeout{** the default language instead.}%
\else
\language=\csname l@#1\endcsname
\fi
#2}}
\providecommand{\BIBdecl}{\relax}
\BIBdecl

\bibitem{ronneberger2015u}
O.~Ronneberger, P.~Fischer, and T.~Brox, ``U-net: Convolutional networks for
  biomedical image segmentation,'' in \emph{International Conference on Medical
  image computing and computer-assisted intervention}.\hskip 1em plus 0.5em
  minus 0.4em\relax Springer, 2015, pp. 234--241.

\bibitem{romano2016raisr}
Y.~Romano, J.~Isidoro, and P.~Milanfar, ``Raisr: rapid and accurate image super
  resolution,'' \emph{IEEE Transactions on Computational Imaging}, vol.~3,
  no.~1, pp. 110--125, 2016.

\bibitem{chen2018efficient}
Y.~Chen, F.~Shi, A.~G. Christodoulou, Y.~Xie, Z.~Zhou, and D.~Li, ``Efficient
  and accurate mri super-resolution using a generative adversarial network and
  3d multi-level densely connected network,'' in \emph{International Conference
  on Medical Image Computing and Computer-Assisted Intervention}.\hskip 1em
  plus 0.5em minus 0.4em\relax Springer, 2018, pp. 91--99.

\bibitem{li2020adaptive}
B.~Li, C.~Wu, M.~Wang, K.~Charan, and C.~Xu, ``An adaptive excitation source
  for high-speed multiphoton microscopy,'' \emph{Nature methods}, vol.~17,
  no.~2, pp. 163--166, 2020.

\bibitem{zhou2020w2s}
R.~Zhou, M.~E. Helou, D.~Sage, T.~Laroche, A.~Seitz, and S.~S{\"u}sstrunk,
  ``W2s: microscopy data with joint denoising and super-resolution for
  widefield to sim mapping,'' in \emph{European Conference on Computer
  Vision}.\hskip 1em plus 0.5em minus 0.4em\relax Springer, 2020, pp. 474--491.

\bibitem{wang2019deep}
H.~Wang, Y.~Rivenson, Y.~Jin, Z.~Wei, R.~Gao, H.~G{\"u}nayd{\i}n, L.~A.
  Bentolila, C.~Kural, and A.~Ozcan, ``Deep learning enables cross-modality
  super-resolution in fluorescence microscopy,'' \emph{Nature methods},
  vol.~16, no.~1, pp. 103--110, 2019.

\bibitem{qiao2021evaluation}
C.~Qiao, D.~Li, Y.~Guo, C.~Liu, T.~Jiang, Q.~Dai, and D.~Li, ``Evaluation and
  development of deep neural networks for image super-resolution in optical
  microscopy,'' \emph{Nature Methods}, vol.~18, no.~2, pp. 194--202, 2021.

\bibitem{hu2019runet}
X.~Hu, M.~A. Naiel, A.~Wong, M.~Lamm, and P.~Fieguth, ``Runet: A robust unet
  architecture for image super-resolution,'' in \emph{Proceedings of the
  IEEE/CVF Conference on Computer Vision and Pattern Recognition Workshops},
  2019, pp. 0--0.

\bibitem{al2018deep}
Y.~Al-Kofahi, A.~Zaltsman, R.~Graves, W.~Marshall, and M.~Rusu, ``A deep
  learning-based algorithm for 2-d cell segmentation in microscopy images,''
  \emph{BMC bioinformatics}, vol.~19, no.~1, pp. 1--11, 2018.

\bibitem{hatipoglu2017cell}
N.~Hatipoglu and G.~Bilgin, ``Cell segmentation in histopathological images
  with deep learning algorithms by utilizing spatial relationships,''
  \emph{Medical \& biological engineering \& computing}, vol.~55, no.~10, pp.
  1829--1848, 2017.

\bibitem{fang2021deep}
L.~Fang, F.~Monroe, S.~W. Novak, L.~Kirk, C.~R. Schiavon, S.~B. Yu, T.~Zhang,
  M.~Wu, K.~Kastner, A.~A. Latif \emph{et~al.}, ``Deep learning-based
  point-scanning super-resolution imaging,'' \emph{Nature methods}, vol.~18,
  no.~4, pp. 406--416, 2021.

\bibitem{edlund2021livecell}
C.~Edlund, T.~R. Jackson, N.~Khalid, N.~Bevan, T.~Dale, A.~Dengel, S.~Ahmed,
  J.~Trygg, and R.~Sj{\"o}gren, ``Livecell—a large-scale dataset for
  label-free live cell segmentation,'' \emph{Nature methods}, vol.~18, no.~9,
  pp. 1038--1045, 2021.

\bibitem{fidon2017generalised}
L.~Fidon, W.~Li, L.~C. Garcia-Peraza-Herrera, J.~Ekanayake, N.~Kitchen,
  S.~Ourselin, and T.~Vercauteren, ``Generalised wasserstein dice score for
  imbalanced multi-class segmentation using holistic convolutional networks,''
  in \emph{International MICCAI brainlesion workshop}.\hskip 1em plus 0.5em
  minus 0.4em\relax Springer, 2017, pp. 64--76.

\bibitem{bertels2019optimizing}
J.~Bertels, T.~Eelbode, M.~Berman, D.~Vandermeulen, F.~Maes, R.~Bisschops, and
  M.~B. Blaschko, ``Optimizing the dice score and jaccard index for medical
  image segmentation: Theory and practice,'' in \emph{International Conference
  on Medical Image Computing and Computer-Assisted Intervention}.\hskip 1em
  plus 0.5em minus 0.4em\relax Springer, 2019, pp. 92--100.

\bibitem{zhang2017brain}
J.~Zhang, X.~Shen, T.~Zhuo, and H.~Zhou, ``Brain tumor segmentation based on
  refined fully convolutional neural networks with a hierarchical dice loss,''
  \emph{arXiv preprint arXiv:1712.09093}, 2017.

\bibitem{ibtehaz2020multiresunet}
N.~Ibtehaz and M.~S. Rahman, ``Multiresunet: Rethinking the u-net architecture
  for multimodal biomedical image segmentation,'' \emph{Neural Networks}, vol.
  121, pp. 74--87, 2020.

\bibitem{zhou2018unet++}
Z.~Zhou, M.~M. Rahman~Siddiquee, N.~Tajbakhsh, and J.~Liang, ``Unet++: A nested
  u-net architecture for medical image segmentation,'' in \emph{Deep learning
  in medical image analysis and multimodal learning for clinical decision
  support}.\hskip 1em plus 0.5em minus 0.4em\relax Springer, 2018, pp. 3--11.

\bibitem{milesi}
A.~Milesi, ``Pytorch-unet,'' \url{https://github.com/milesial/Pytorch-UNet},
  2022.

\end{thebibliography}
\bibliographystyle{IEEEtran}  

\end{document}